\documentclass[preprintnumbers,floatfix,12pt,superscriptaddress,nofootinbib]{revtex4}

\usepackage{amsfonts}
\usepackage{amssymb,epsfig}
\usepackage{latexsym}
\usepackage{amsmath}
\usepackage{graphicx}

\begin{document}

\title{Polytropic scalar field models of dark energy}
\author{M. Malekjani \footnote{Email: \text{malekjani@basu.ac.ir}}}

 \affiliation{Department of Physics, Faculty of Science, Bu-Ali Sina University, Hamedan 65178,
Iran\\} \affiliation{Research Institute for Astronomy and
Astrophysics of Maragha (RIAAM), Maragha, Iran\\}

\begin{abstract}
\vspace*{1.5cm} \centerline{\bf Abstract} \vspace*{1cm} In this work
we investigate the polytropic gas dark energy model in the non flat
universe. We first calculate the evolution of EoS parameter of the
model as well as the cosmological evolution of Hubble parameter in
the context of polytropic gas dark energy model. Then we reconstruct
the dynamics and the potential of the tachyon and K-essence scalar
field models according to the evolutionary behavior of polytropic
gas model.

\end{abstract}
\maketitle

\newpage
\section{Introduction}
Nowadays our belief is that the current universe is in accelerating
expansion. The results of cosmological experiments: SNe Ia
\cite{c1}, WMAP \cite{c2}, SDSS \cite{c3} and X-ray \cite{c4}
 , have provided the main evidences for this cosmic
acceleration. In the framework of standard cosmology, a new energy
with negative pressure, namely dark energy (DE), is needed to
explain this acceleration.  The cosmological constant with the time
- independent equation of state $w_{\Lambda}=-1$ is the earliest and
simplest candidate of dark energy. Although the cosmological
constant is consistent with observational data, but from the
theoretical viewpoint is faces with the fine-tuning and cosmic
coincidence problems \cite{fin}. In addition to cosmological
constant, the other dynamical dark energy models with time-varying
equation of state have been suggested to explain the cosmic
acceleration. Recent SNe Ia observational data show that the
dynamical dark energy models have a better fit compare with
cosmological constant \cite{c44}. The scalar field models such as
quintessence \cite{c5}, phantom \cite{c6}, quintom \cite{c7},
K-essence \cite{c8}, tachyon \cite{c9} and dilaton \cite{c10}
together with interacting dark energy models such as holographic
\cite{c11} and agegraphic \cite{c12} models are the examples of
dynamical dark energy models.\\ The holographic dark energy model
comes from the holographic principle of quantum gravity \cite{c13}
and the agegraphic model has been proposed based on the uncertainty
relation of quantum mechanics together with general relativity
\cite{c14}.\\
In this work, we focus on the polytropic gas model as a dark energy
model to explain the cosmic acceleration. In stellar astrophysics,
the polytropic gas model can explain the equation of state of
degenerate white dwarfs, neutron stars and also the equation of
state of main sequence stars \cite{c19}. The idea of dark energy
with polytropic gas equation of state has been investigated by U.
Mukhopadhyay and S. Ray in cosmology \cite{ray05}. The  polytropic
gas is a phenomenological model of dark energy. In a
phenomenological model, the pressure $p$ is a function of energy
density $\rho$, i.e., $p=-\rho-f(\rho)$ \cite{c177}. For
$f(\rho)=0$, the equation of state of phenomenological models can
cross $w=-1$, i.e., the cosmological constant model. Nojiri, et al.
investigated four types singularities for some illustrative examples
of phenomenological models \cite{c177}. The polytropic gas model has
a type III. singularity in which the singularity takes place at a
 characteristic scale factor $a_s$.\\
Recently, Karami et al. investigated the interaction between dark
energy and dark matter in polytropic gas scenario, the phantom
behavior of polytropic gas, reconstruction of $f(T)$- gravity from
the polytropic gas and the correspondence between polytropic gas and
agegraphic dark energy model \cite{c17,c18,karam20}. The
cosmological implications of polytropic gas dark energy model is
also discussed in \cite{malek1}. The evolution of deceleration
parameter in the context of polytropic gas dark energy model
represents the decelerated expansion at the early universe and
accelerated phase later as expected. The polytropic gas model has also been studied
from the viewpoint of statefinder analysis in \cite{malek_state}.\\
On the other hands, as we know, the scalar field models are the
effective description of an underlying theory of dark energy. Scalar
fields naturally arise in particle physics including supersymmetric
field theories and string/M theory. The scalar field can reveal the
dynamic and the nature of dark energy. However, the fundamental
theories such as string/M theory do not predict their potential
$V(\phi)$ uniquely. Consequently, it is meaningful to reconstruct
the potential of dark energy model so that these scalar fields can
describe the evolutionary behavior of dark energy model possessing
some significant features of the quantum gravity theory, such as
holographic and agegraphic dark energy models. In this direction,
many works have been done \cite{scalar}. In this paper we
reconstruct the dynamics and the potential of tachyon and the
K-essence scalar fields model according to the evolution of
polytropic gas model.
\section{FRW cosmology and polytropic gas dark energy}
Let us start with non-flat Friedmann-Robertson-Walker (FRW) universe
containing dark energy and dark matter, the corresponding Friedmann
equation is as follows
\begin{equation}\label{frid1}
H^{2}+\frac{k}{a^{2}}=\frac{1}{3M_{p}^{2}}(\rho _{m}+\rho _{d})
\end{equation}%
where $H$ is the Hubble parameter, $M_p$ is the reduced Plank mass
and $k=1,0,-1$ is a curvature parameter corresponding to a closed,
flat and open universe, respectively. $\rho_m$ and $\rho_{\Lambda}$
are the energy density of dark matter and dark energy, respectively.
Recent observations support a closed universe with a tiny positive
small curvature $\Omega _{k}=\simeq0.02$ \cite{c20}.\\
In the case of dimensionless energy densities
\begin{equation}\label{denergy}
\Omega_{m}=\frac{\rho_m}{\rho_c}=\frac{\rho_m}{3M_p^2H^2}, ~~~\\
\Omega_{d}=\frac{\rho_{d}}{\rho_c}=\frac{\rho_{d}}{3M_p^2H^2}~~\\
\Omega_k=\frac{k}{a^2H^2},
\end{equation}
 the Friedmann equation (\ref{frid1}) can be written as
\begin{equation}
\Omega _{m}+\Omega _{d}=1+\Omega _{k}.  \label{Freq2}
\end{equation}%
The conservation equations for dark matter and dark energy are given
by
\begin{eqnarray}
\dot{\rho _{m}}+3H\rho _{m}=0, \label{contm}\\
\dot{\rho _{\Lambda}}+3H(\rho_{d}+p_{d})=0, \label{contd}
\end{eqnarray}%
The equation of state (EoS) of polytropic gas is given by
\begin{equation}\label{poly}
p_{d}=K\rho_{d}^{1+\frac{1}{n}},
\end{equation}
where $K$ and $n$ are constants of the model \cite{c19}. Inserting
Eq.(\ref{poly}) in (\ref{contd}) and integrating obtains the energy
density of polytropic gas dark energy as
 \begin{equation}\label{rho1}
 \rho_{d}=\left(\frac{1}{Ba^{3/n}-K}\right)^n,
 \end{equation}
where $B$ is the integration constant and $a$ is the scale factor.
For $Ba^{3/n}>K$ the energy density of polytropic gas is positive
for any odd and event number of $n$. But in the case of $Ba^{3/n}<K$
the energy density is positive only for even numbers. The phantom
behavior of interacting polytropic gas dark energy has been
calculated in \cite{c17}. The phenomenological equation of state
such as chaplygin gas and polytropic gas models usually suffers from
the singularity problem in which the energy density tends to
infinity. The singularity of dark energy models has been discussed
in \cite{noji12}. In the case of $Ba^{3/n}=K$, we have
$\rho_d\rightarrow \infty$ and the polytropic gas has a finite-time
singularity at $a_c=(K/B)^{n/3}$. This type of singularity has been
named by type III singularity \cite{noji12}.\\
Tacking the time derivative of (\ref{rho1}) with respect to time
obtains
\begin{equation}\label{dotrho}
\dot{\rho_{d}}=-3BHa^{\frac{3}{n}}\rho_{d}^{1+\frac{1}{n}}
\end{equation}

Substituting (\ref{dotrho})in conservation equation of dark energy
component (\ref{contd}) and using (\ref{rho1}),
$p_{d}=w_{d}\rho_{d}$, in (\ref{contd}) we obtain the EoS parameter
of polytropic gas dark energy model as
\begin{equation}\label{eos1}
 w_{d}=-1-\frac{a^{\frac{3}{n}}}{c-a^{\frac{3}{n}}}
\end{equation}
where $c=K/B$. Here one can see that the polytropic gas model can
cross the phantom line, i.e. $w_{d}=-1$, when $c>a^{3/n}$. Also at
the early time ($a\rightarrow 0$), the polytropic gas mimics
the cosmological constant, i.e. $w_{d}\rightarrow -1$.\\
In figure (1), the evolution of EoS parameter $w_{d}$ is plotted as
a function of redshift parameter $z$. Note that the redshift
parameter is related to scale factor by $a=1/(1+z)$. Here we
conclude that the polytropic gas model for the selected model
parameter:$c=2$ and even numbers: \{$n=2,4,6$\} behaves as a phantom
dark energy as indicated in left panel. Also in the right panel the
phantom regime can be obtained for different illustrative values
\{$c=2,3,4$\} and even number $n=2$.\\
From (\ref{rho1}), we see that the polytropic gas has a singularity
at $a_s=c^{n/3}$. For $c<1$, this singularity takes place at $a_s<1$
(past). For $c=1$ the singularity occurs at the present time $a_s=1$
and in the case of $c>1$ it occurs at future, i.e. $a_s>1$. In
figure (2) we show the singularity of polytropic gas model for
different values of $c$. In upper left panel we choose $c=2$. In
this case the singularity of the model tacks place at future
($a_s=1.58$). In upper right panel $c=1$ and the singularity occurs
at the present time $a_s=1$. Eventually at the lower panel, for
$c=0.4$, the singularity occurs at the past time $a_s=0.57$. One of
the advantage of polytropic gas model is that this model can behaves
as a phantom dark energy without a need to interaction between dark
matter and dark energy. From figure (1), we see the that the phantom
regime ($w_d<-1$) for polytropic gas as indicated by (\ref{eos1}).
But the other theoretical models of dark energy such as holographic
and agegraphic can not enter the phantom regime without interaction
term, for example see \cite{cai22,setare22}. From (\ref{eos1}), we
also see that for the condition of $c<a^{3/n}$, the polytropic gas
can behave as a quintessence model,i.e., $-1<w_d<0$. The problem of
phenomenological models such as polytropic gas model is that,
because of singularity at $a_s$, the cosmology for these models can
be defined only in the interval $0<a<a_s$, i.e., from the Big Bang
epoch to the singularity epoch at $a_s$. In other word, the
polytropic gas model can describe the acceleration of the universe
from the Big Bang epoch up to singularity epoch at the scale factor
$a_s$. In the next section we reconstruct the potential and the
dynamics of tachyon scalar field according to the evolution of
phantom polytropic gas dark energy.\\
We now obtain the Hubble parameter in the context of polytropic dark
energy model. From the conservation equations (\ref{contm}),
(\ref{contd}) and using the dimensionless energy densities in
(\ref{denergy}), we have
\begin{eqnarray}\label{denergyd}
\rho_m=\rho_{m0}a^{-3}\\
\rho_d=\rho_{d0}a^{-3[1+w_d(a)]}\label{denergydd}
\end{eqnarray}

Inserting (\ref{denergyd},\ref{denergydd}) in Friedmann equation
(\ref{frid1}) and using the dimensionless energy densities
(\ref{denergy}), we obtain the Hubble parameter as
\begin{equation}\label{hubbb}
H(a)=H_0\sqrt{\Omega_{m0}a^{-3}+\Omega_{d0}a^{-3[1+w_d(a)]}-\Omega_{k0}a^{-2}}
\end{equation}
where $w_d(a)$ is given by (\ref{eos1}).\\
 In figure (3), we plot the
evolution of dimensionless Hubble parameter, $E(a)=H(a)/H_0$, for
polytropic gas model. In left panel, we fix $c$ and vary the
parameter $n$. The smaller value the parameter $n$ is taken, the
bigger the Hubble parameter expansion rate E(a) can reach. In right
panel, by fixing $n$, we vary the parameter $c$. The dimensionless
Hubble parameter E(a) is bigger for larger value of $c$. One can
explicitly see that both the model parameters $n$ and $c$ can impact
the expansion of the universe.

\newpage
\begin{center}
\begin{figure}[!htb]
\includegraphics[width=7.6cm]{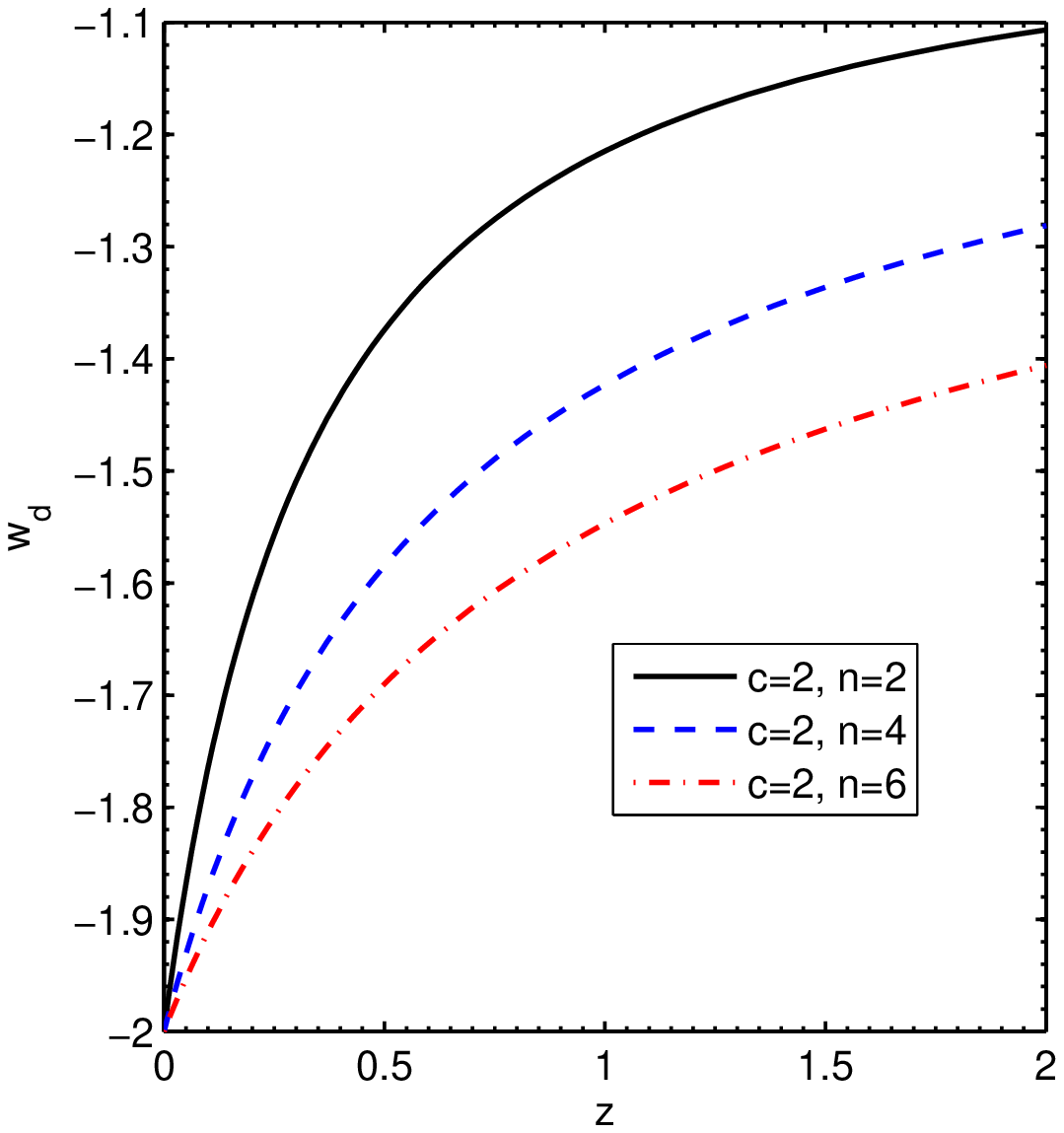}\includegraphics[width=7cm]{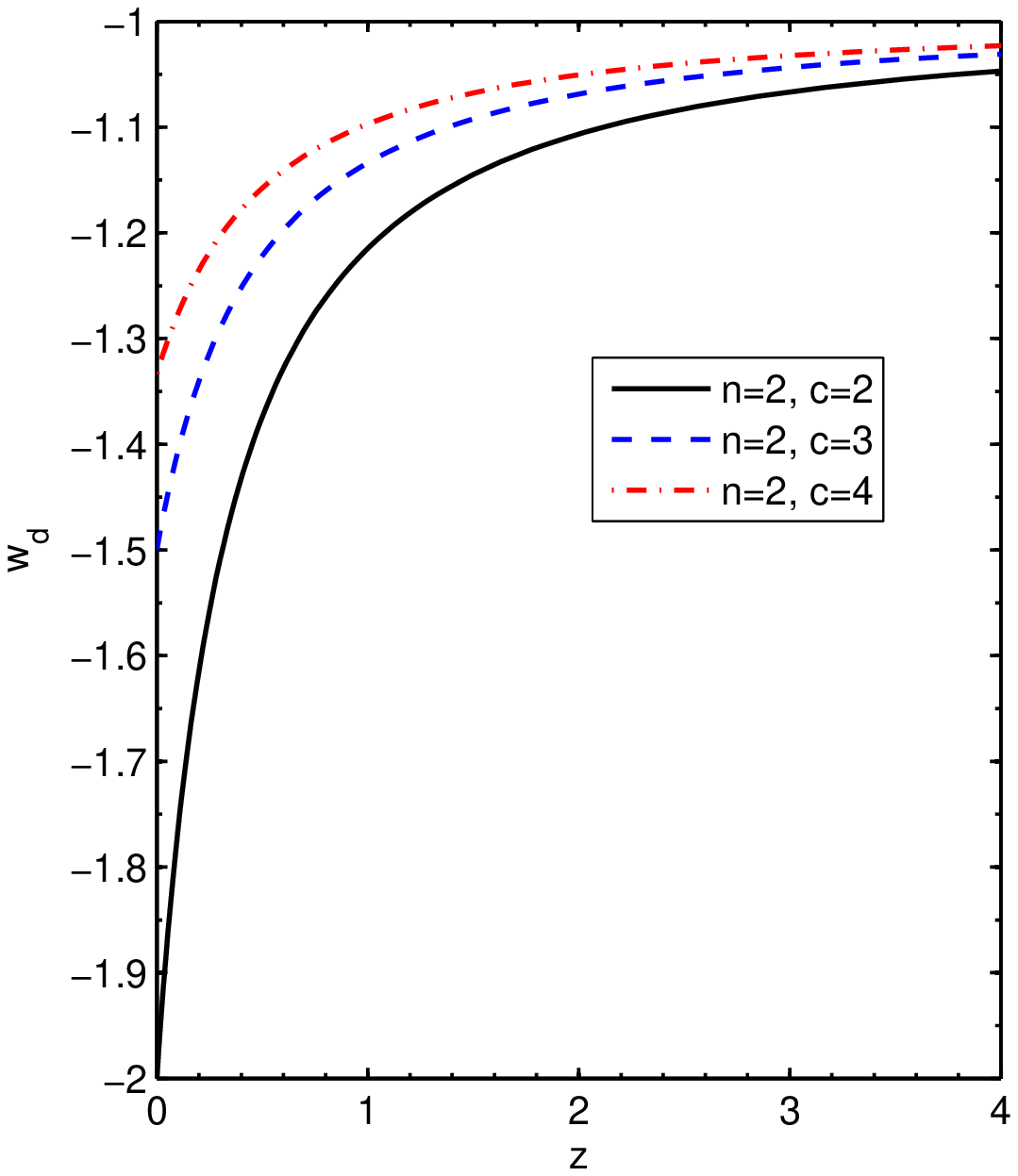}
\caption{The EoS parameter of polytropic gas model for different
values of model parameters: $n$ and $c$ as described in legend.}
\end{figure}
\end{center}

\newpage
\begin{center}
\begin{figure}[!htb]
\includegraphics[width=7.6cm]{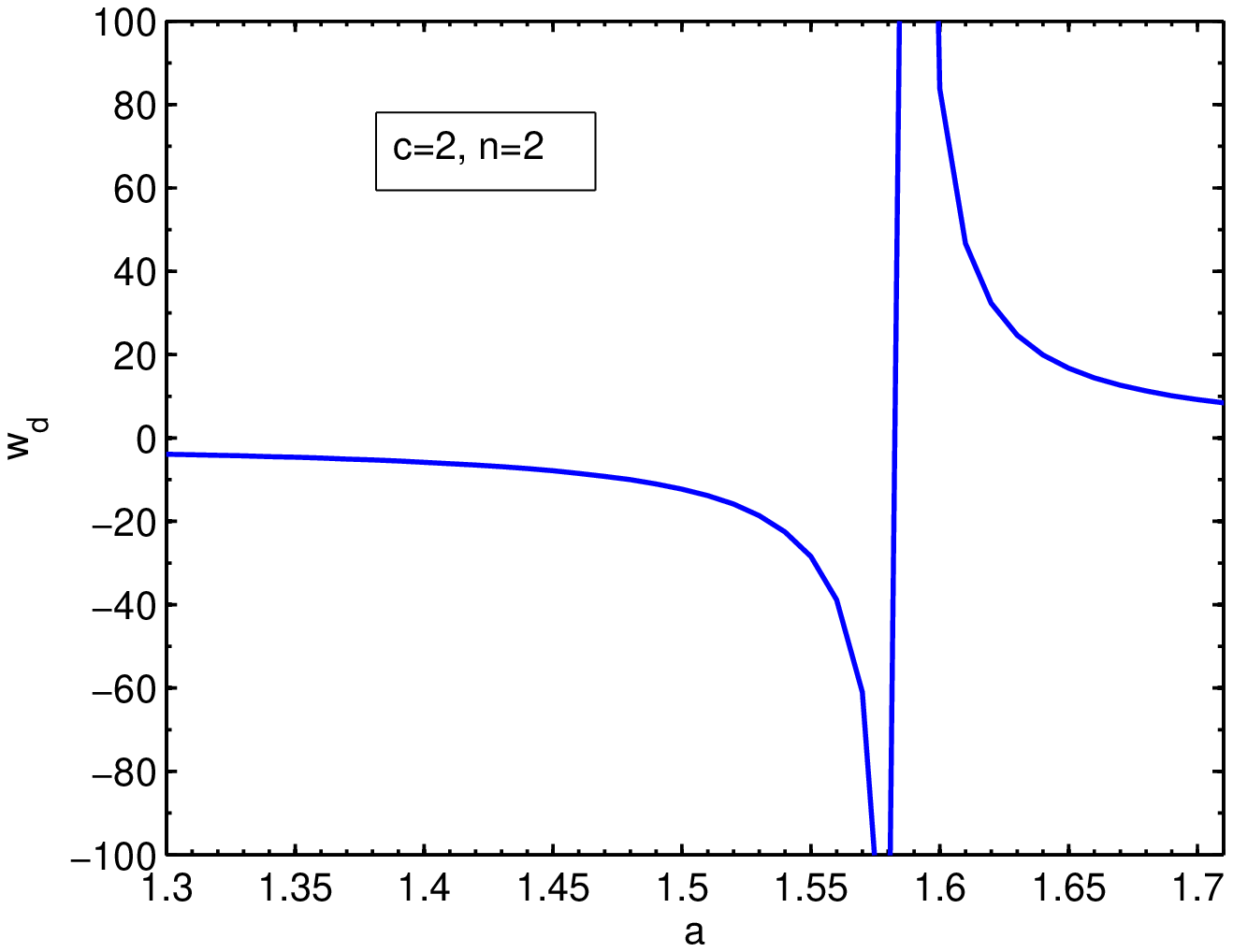}\includegraphics[width=7.6cm]{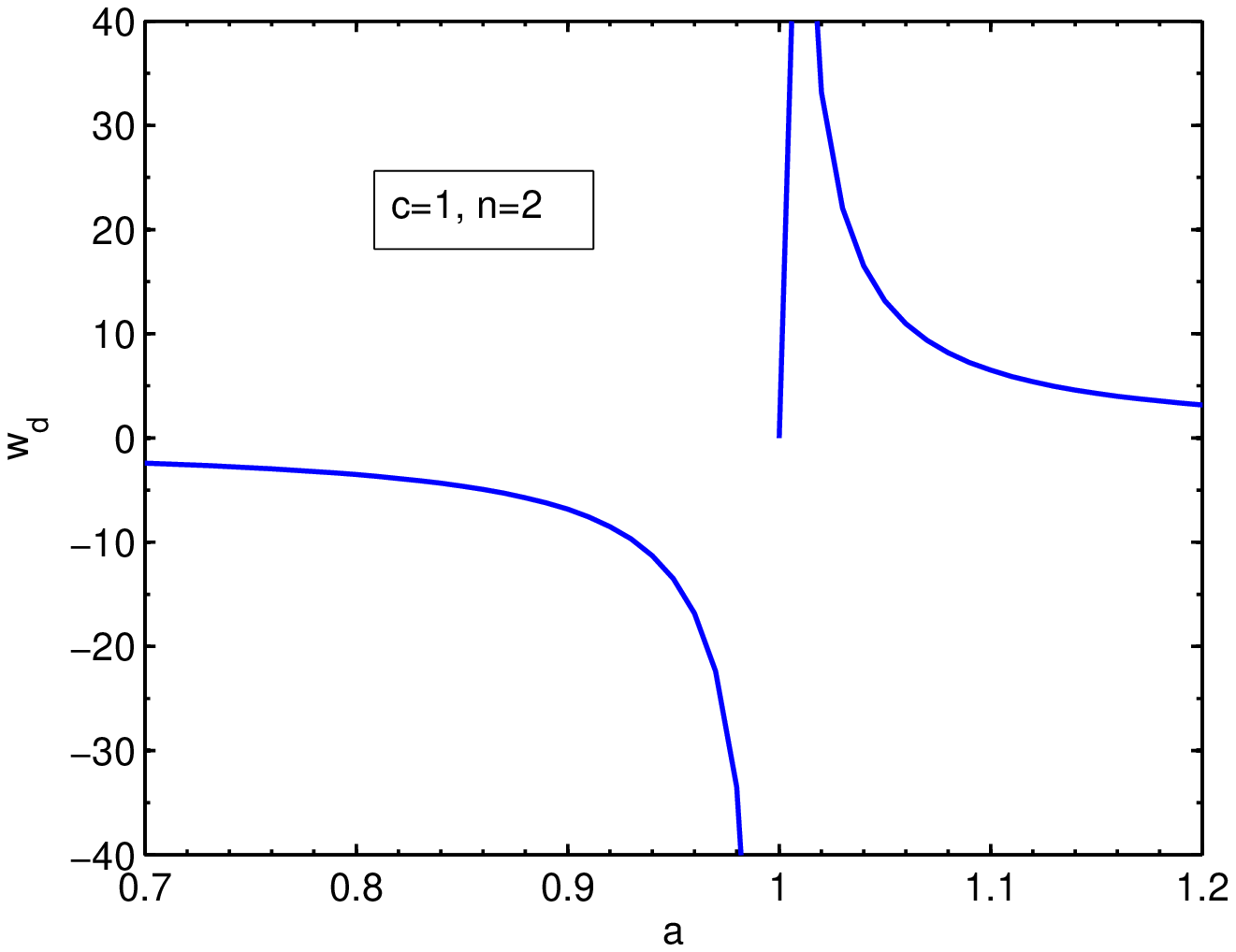}
\includegraphics[width=7.6cm]{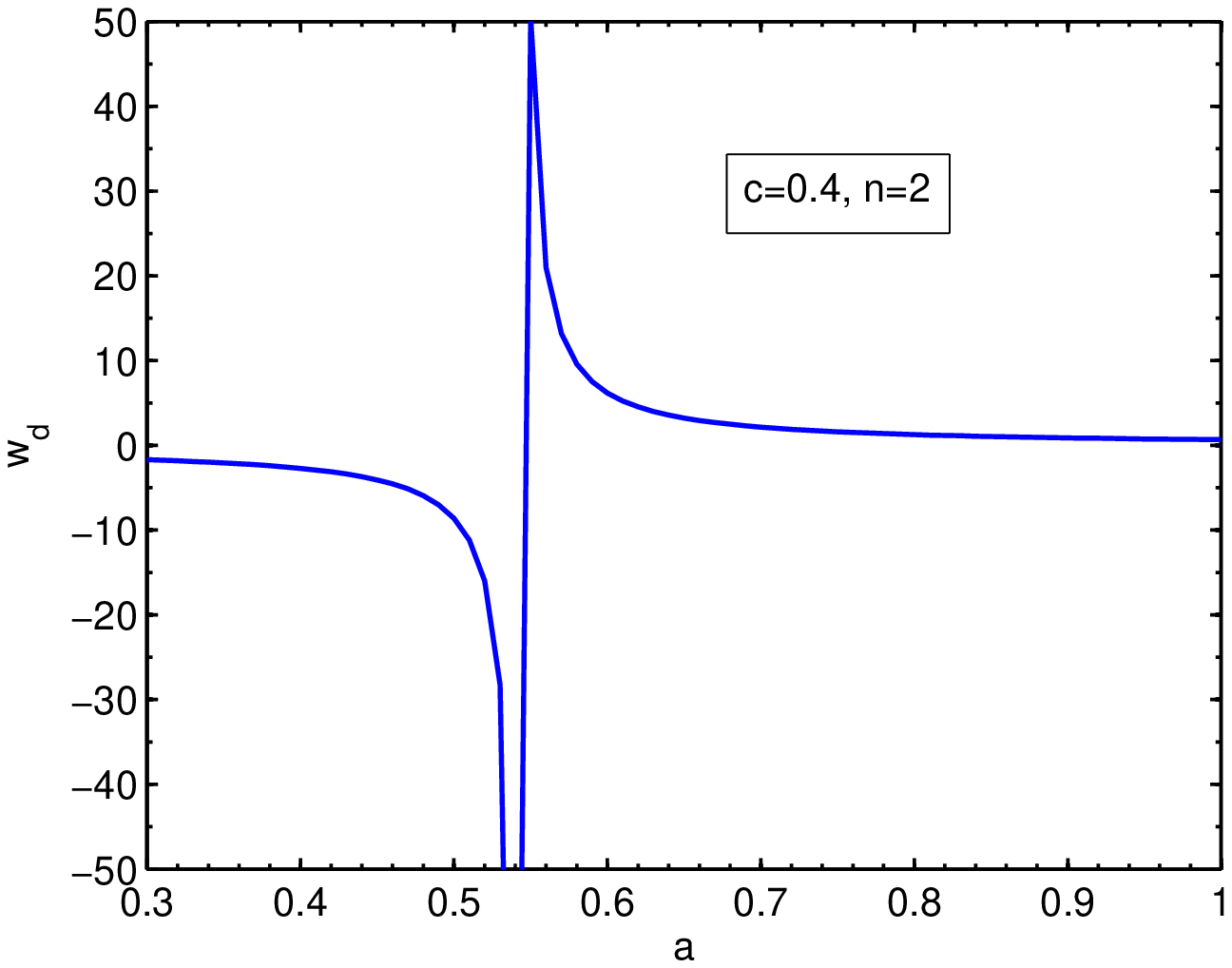}
\caption{The singularity of polytropic gas model for different
values of model parameter $c$.}
\end{figure}
\end{center}

\newpage
\begin{center}
\begin{figure}[!htb]
\includegraphics[width=7.6cm]{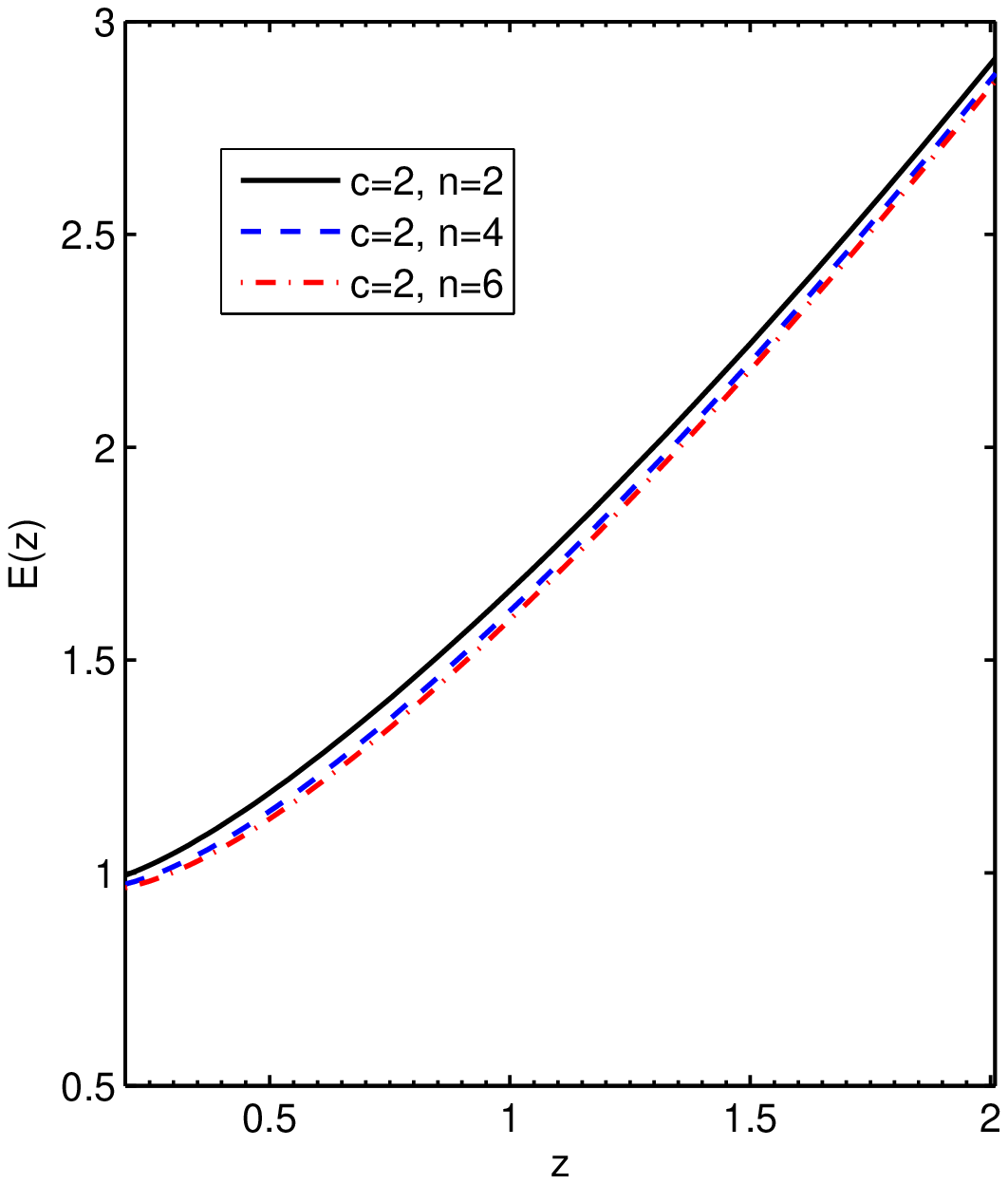}\includegraphics[width=7.8cm]{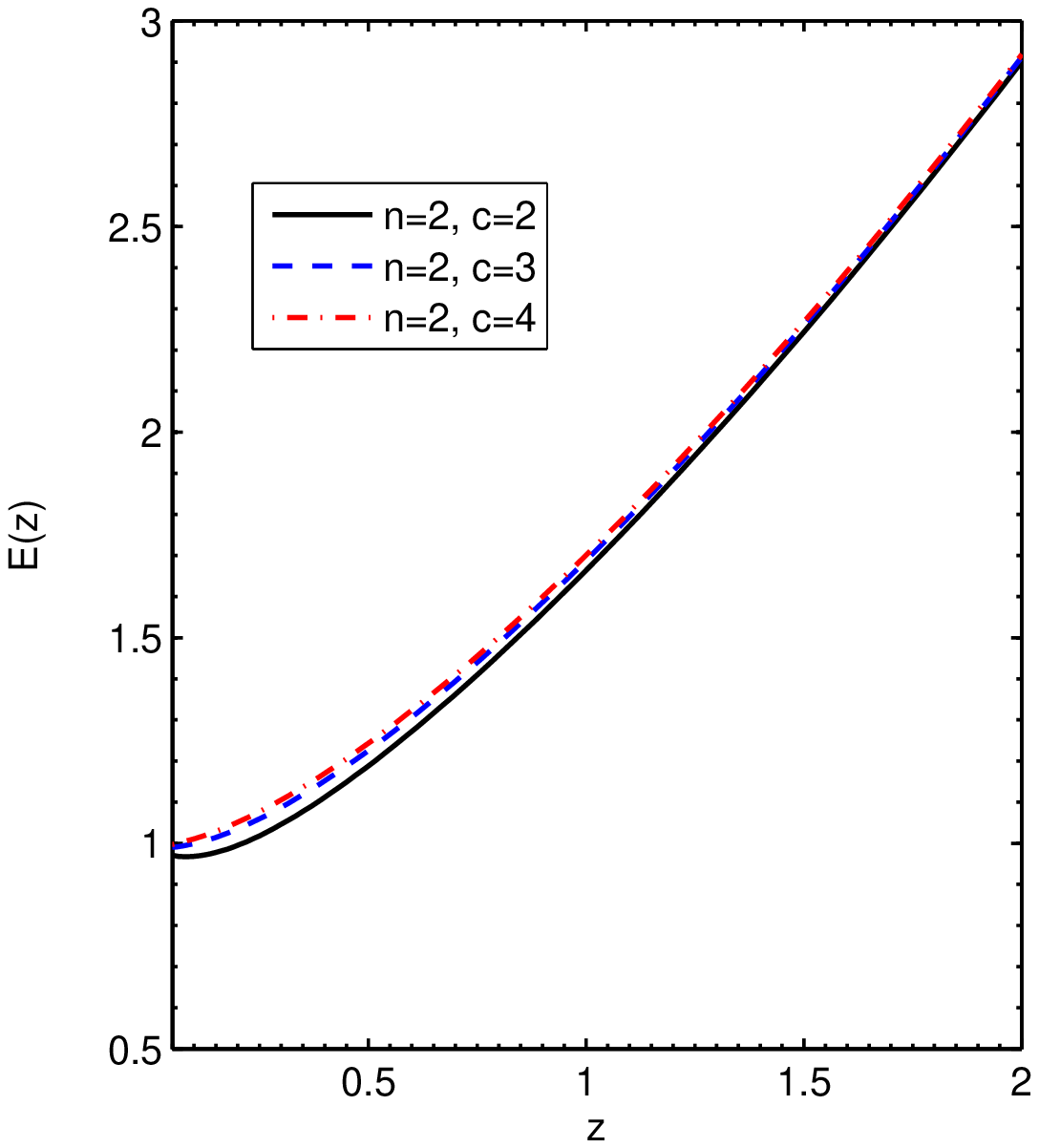}
\caption{The evolution of dimensionless Hubble parameter as a
function of redshift parameter $z$ for different values of $n$ and
$c$ as described in legend.}
\end{figure}
\end{center}

\newpage

\section{Tachyon reconstruction of polytropic gas model}
Here we establish a correspondence between the polytropic gas model
with the tachyon scalar field. We reconstruct the potential and the
dynamics of tachyon field according to the evolution of polytropic gas model.\\
The tachyon scalar field can be considered as a source of dark
energy \cite{c24}. The tachyon is an unstable field which can be
used in string theory through its role in the Dirac-Born-Infeld
(DBI) action to describe the D-bran action \cite{c25}. The effective
Lagrangian for the tachyon field is given by
\[
\mathcal{L}=-V(\phi )\sqrt{1-g^{\mu \nu }\partial _{\mu }\phi
\partial _{\nu }\phi },
\]%
where $V(\phi )$ is the potential of tachyon field. The energy
density and pressure of tachyon field are given by  \cite{c25}
\begin{equation}
\rho _{\phi }=\frac{V(\phi )}{\sqrt{1-\dot{\phi}^{2}}},
\label{tach1}
\end{equation}%
\begin{equation}
p_{\phi }=-V(\phi )\sqrt{1-\dot{\phi}^{2}}.
\end{equation}%
The EoS parameter of tachyon field can be given by
\begin{equation}
w_{\phi }=\frac{p_{\phi }}{\rho _{\phi }}=\dot{\phi}^{2}-1.
\label{eos_tach}
\end{equation}%
From (\ref{tach1}), we see that in the case of $-1<\dot{\phi}<1$ or
in the case of $\dot{\phi}^2<0$ tachyon field has a real energy
density. Consequently, from (\ref{eos_tach}), it is clear to see
that in the first case the EoS parameter of tachyon is constrained
to $-1<w_{\phi}<0$ and therefor the tachyon field can interpret the
accelerates expansion of universe, but can not enter the phantom
regime, i.e. $w_{d}<-1$. In the later case, $\dot{\phi}^2<0$, we see
$w_{\phi}<-1$, and
the phantom regime can be crossed by tachyon.\\
By equating the relations (\ref{eos1}) and (\ref{eos_tach}) and also
(\ref{rho1}) with (\ref{tach1}), we reconstruct the potential and
the dynamics of tachyon according to evolution of interacting
polytropic gas model as follows

\begin{equation}
w_{d}=-1-\frac{a^{\frac{3}{n}}}{c-a^{\frac{3}{n}}}=\dot{\phi}^{2}-1.
\end{equation}

\begin{equation}
\rho_{d}=\left(\frac{1}{Ba^{\frac{3}{n}}-K}\right)^n=\frac{V(\phi
)}{\sqrt{1-\dot{\phi}^{2}}}
\end{equation}

Therefor we obtain the following expressions for the dynamics and
potential of tachyon field
\begin{equation}\label{dotphi1}
\dot{\phi}^{2}=-\frac{a^{\frac{3}{n}}}{c-a^{\frac{3}{n}}}
\end{equation}

\begin{equation}\label{poten11}
V(\phi)=\sqrt{1+\frac{a^{\frac{3}{n}}}{c-a^{\frac{3}{n}}}}
\left(\frac{1}{Ba^{\frac{3}{n}}-K}\right)^n
\end{equation}

For $c>a^{3/n}$, from (\ref{dotphi1}), we obtain $\dot{\phi}^{2}<0$
which represents the phantom behavior of tachyon field.
 By definition $\phi=i\psi$ and changing the time derivative
to the derivative with respect to logarithmic scale factor, i.e.
$d/dt=H d/dx$, the scalar field $\psi$ can be integrated from
(\ref{dotphi1}) as follows
\begin{equation}\label{dynam11}
\psi(a)-\psi(a_0)=\int_0^a{\frac{1}{aH(a)}\sqrt{\frac{a^{\frac{3}{n}}}{c-a^{\frac{3}{n}}}}}
da
\end{equation}
where $x=\ln{a}$ and $H(a)$ is given by (\ref{hubbb}).
Here we assume the present value of scale factor as $a_0=1$.\\
The potential and the dynamics of reconstructed tachyon field
according to the evolution of polytropic dark energy are given by
relations (\ref{poten11}) and (\ref{dynam11}), respectively.
 Unfortunately, due to the complexity of the equations
involved, the above relations cannot be integrated analytically.
Hence we should use the numerical method to calculate the above
integrations. In figure (4), we show the evolution of the scalar
field for different values of the model parameters $n$ and $c$ as a
function of redshift parameter $z=1/a-1$. Here, for simplicity, we
choose $\psi(z=0)=0$. We can explicitly see the dynamics of the
scalar field where the scalar field decreases from up to zero at the
present time. In left panel we find a faster rate of evolution when
$n$ increases. Also from the right panel we see the faster evolution
of dynamics of reconstructed tachyon field for lower values of model
parameter $c$. In figure (5), the reconstructed tachyon potential
$V(\phi)$ is plotted for different values of model parameter $n$ and
$c$. Here we see that the reconstructed potential $V(\psi)$ has a
nonzero minima  at the early stage of universe ($z\geq5$) which
indicate the cosmological constant behavior of the model in the past
time.From the left and right panels, we see the faster evolution of
potential for larger values of $n$ and smaller values of $c$,
respectively.
\newpage
\begin{center}
\begin{figure}[!htb]
\includegraphics[width=7.6cm]{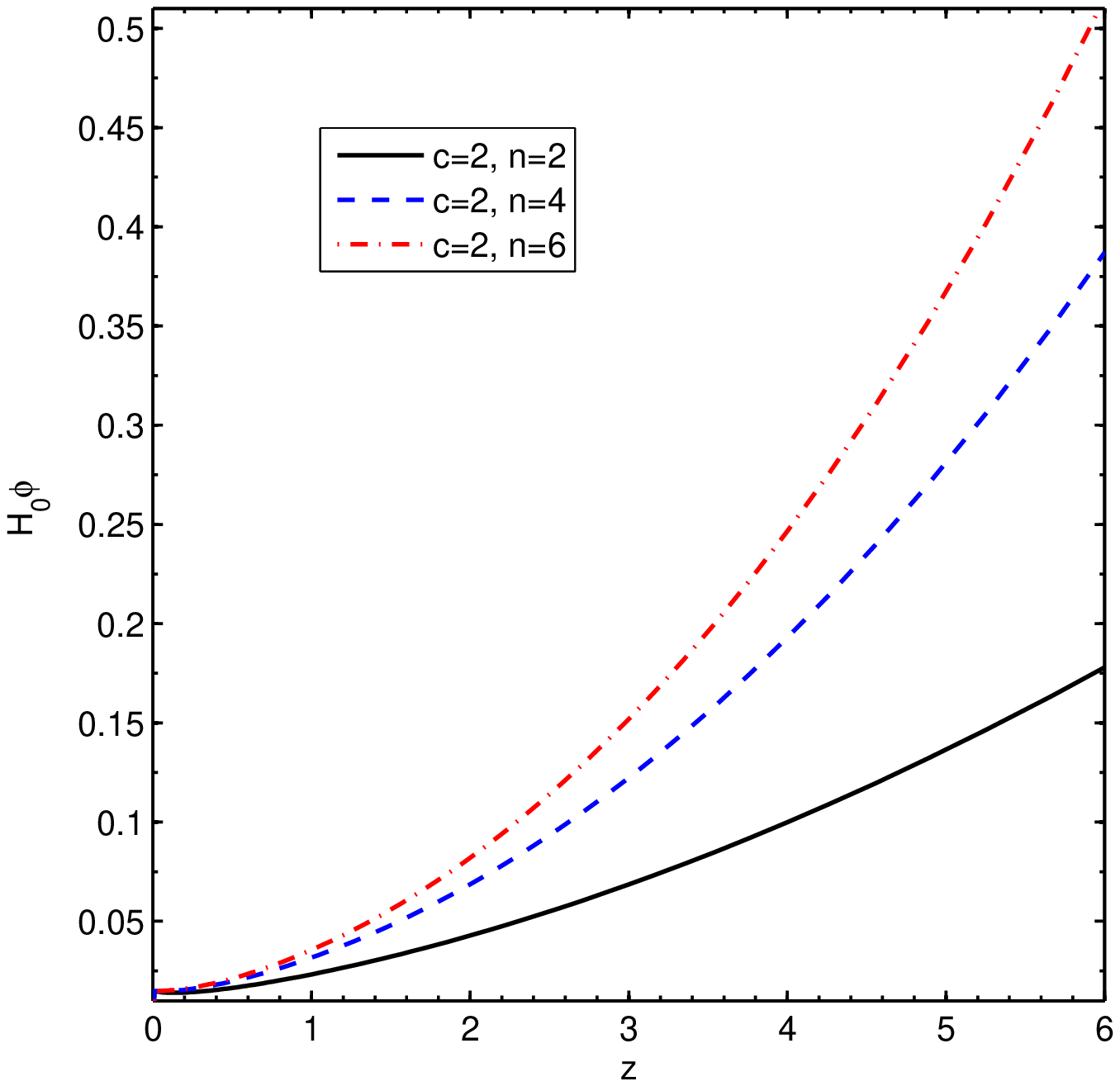}\includegraphics[width=7.6cm]{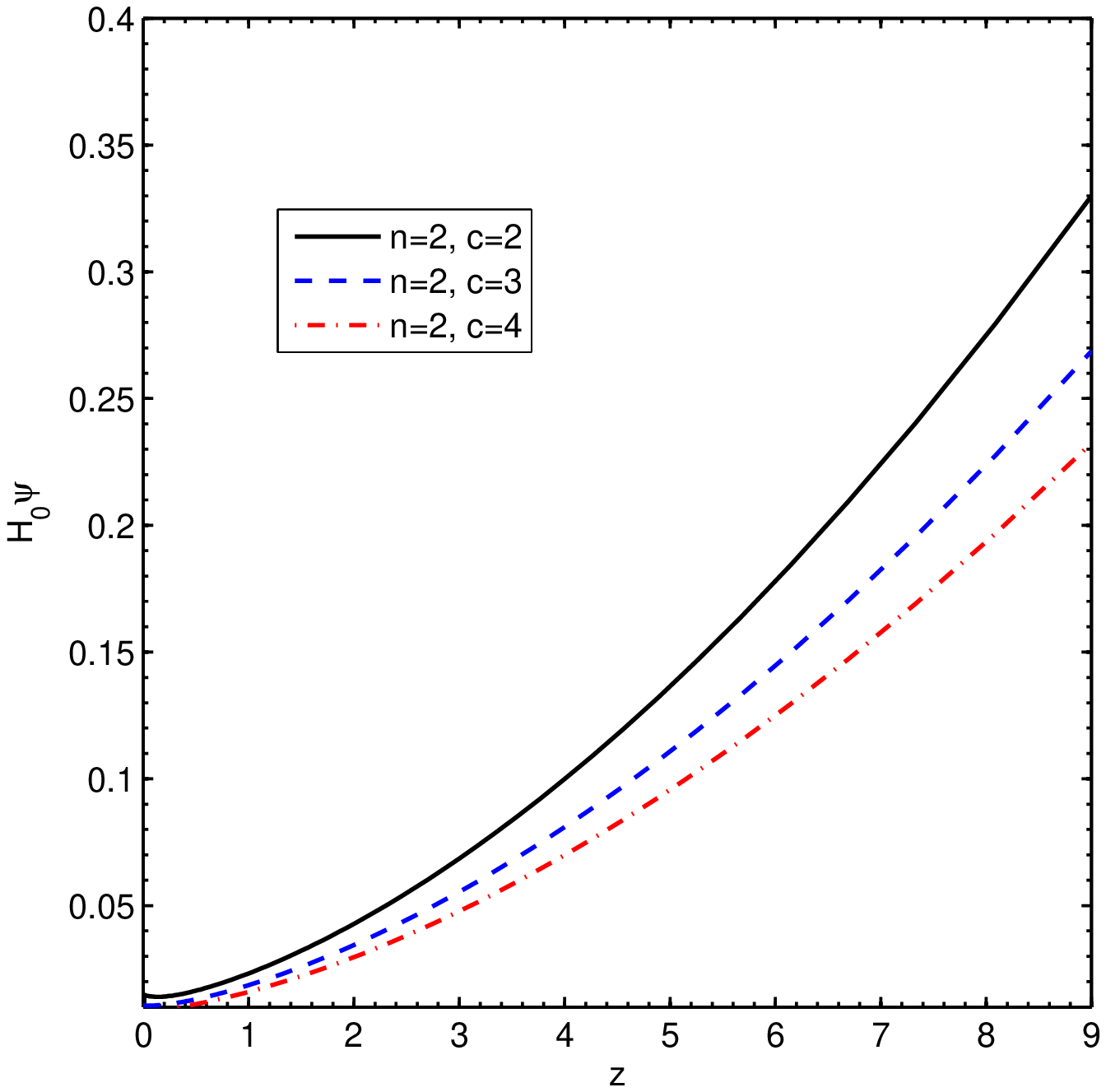}
\caption{The evolution of scalar field $\psi$ in terms of
cosmological redshift parameter $z$ for different values of $n$ and
$c$ as described in legend.}
\end{figure}
\end{center}

\begin{center}
\begin{figure}[!htb]
\includegraphics[width=7.6cm]{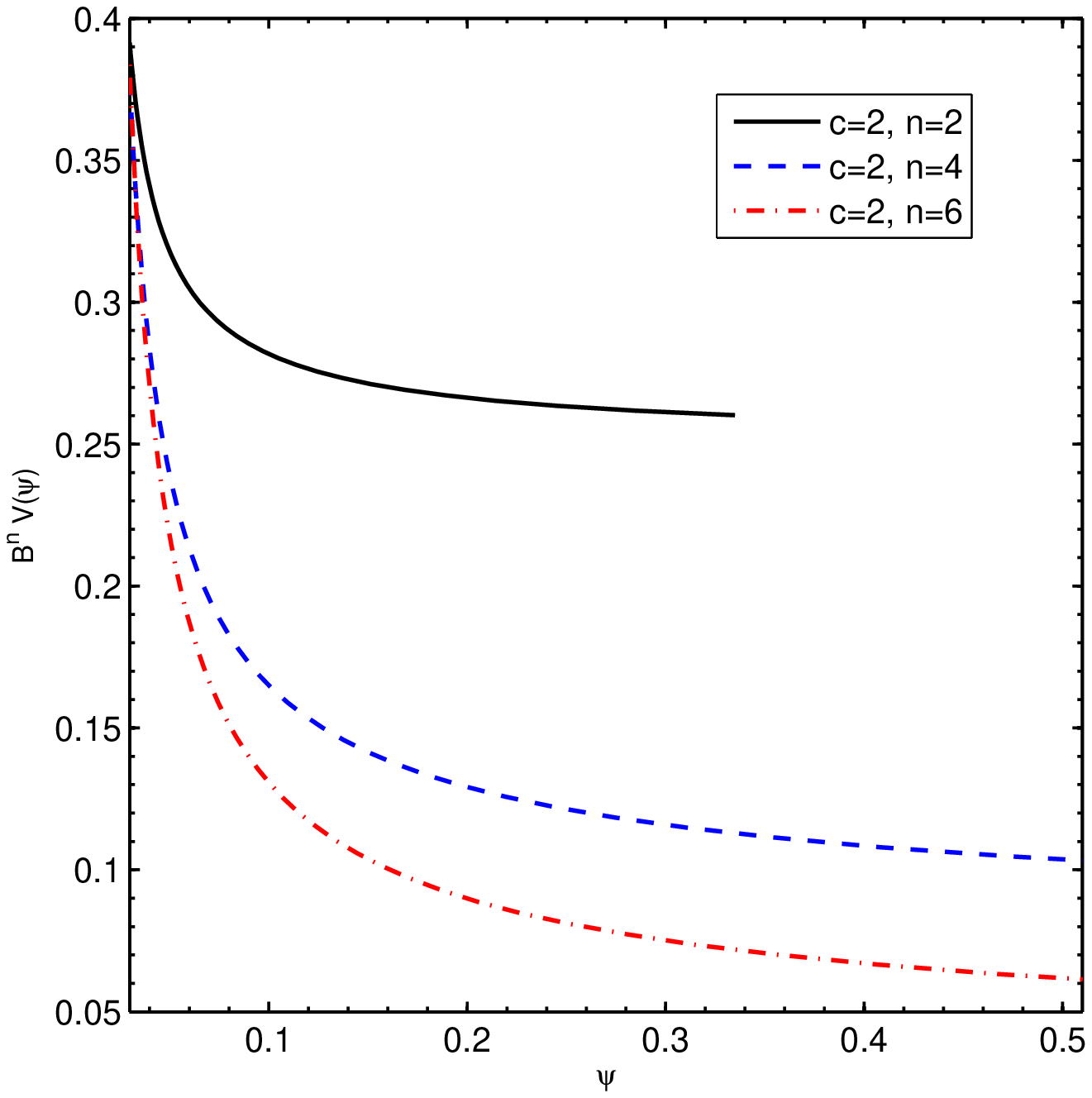}\includegraphics[width=7.6cm]{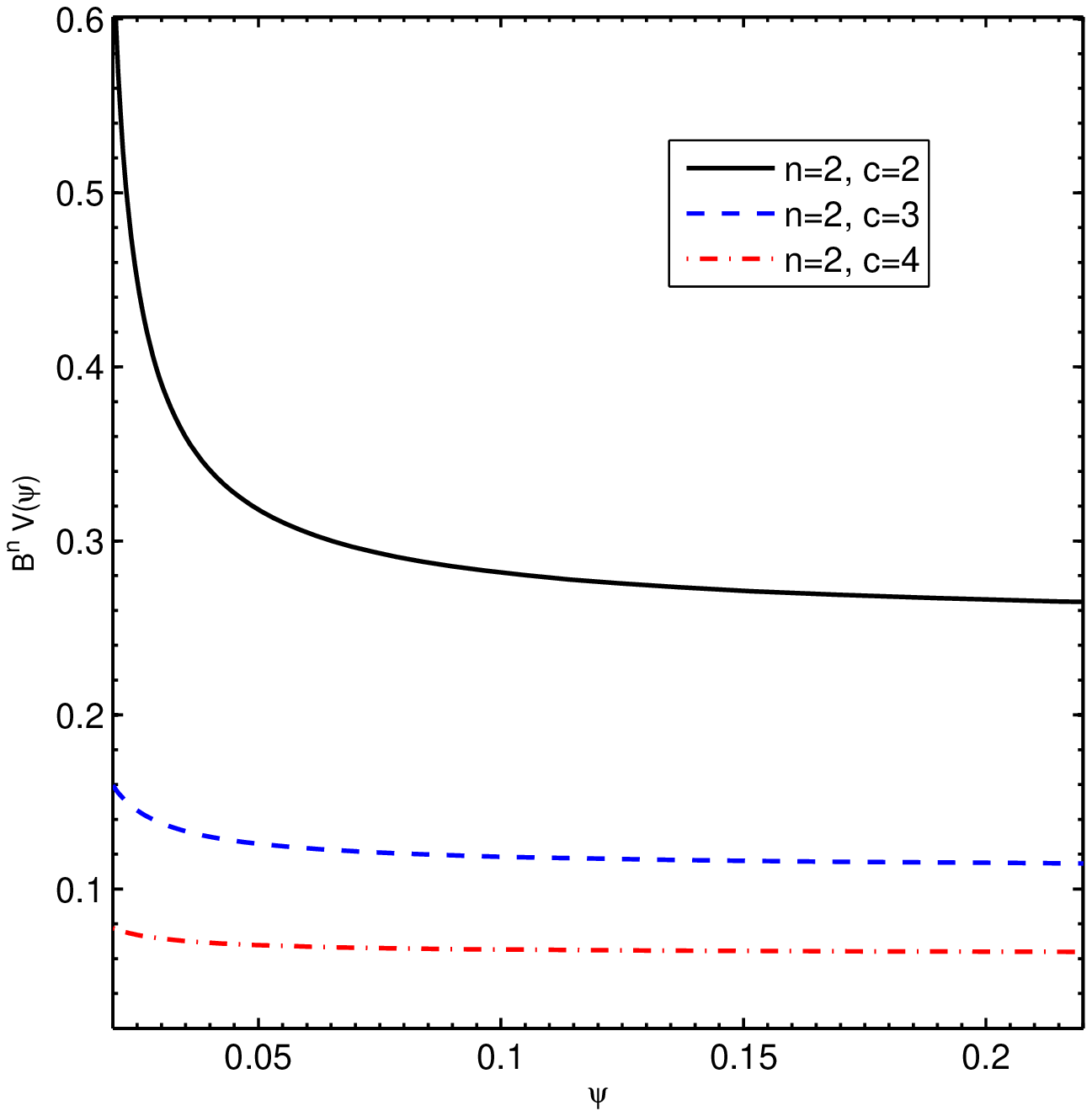}
\caption{The reconstructed potential $V(\psi)$ for different values
of model parameters of tachyon polytropic gas dark energy model as
described in legend.}
\end{figure}
\end{center}

\section{K-essence reconstruction of polytropic gas model}
The propose of the K-essence scalar field was motivated from the
Born-Infeld action of string theory. This kind of scalar field can
interpret the late time acceleration of the universe \cite{c26}. The
K-essence scalar field is given by following action \cite{c27}:
\begin{equation}
S=\int d^{4}x\sqrt{-g}\text{ }p(\phi ,\chi ),
\end{equation}%
where the Lagrangian density $p(\phi ,\chi )$ corresponds to the
pressure density and energy density via the following equations:
\begin{equation}
p(\phi ,\chi )=f(\phi )(-\chi +\chi ^{2}),
\end{equation}%
\begin{equation}
\rho (\phi ,\chi )=f(\phi )(-\chi +3\chi ^{2}).
\end{equation}%
Therefore, the EoS parameter of K-essence can be obtained as follows
\begin{equation}
\omega _{K}=\frac{p(\phi ,\chi )}{\rho (\phi ,\chi )}=\frac{\chi -1}{3\chi -1%
}.  \label{w_k}
\end{equation}%
By equating(\ref{eos1}) and (\ref{w_k}), we have
\begin{equation}
w_{d}=-1-\frac{a^{\frac{3}{n}}}{c-a^{\frac{3}{n}}}=\frac{\chi-1}{3\chi-1}
\end{equation}
Therefore the parameter $\chi$ can be obtained as
\begin{equation}\label{poten77}
\chi=\frac{2+\frac{a^{\frac{3}{n}}}{c-a^{\frac{3}{n}}}}
{4+3\frac{a^{\frac{3}{n}}}{c-a^{\frac{3}{n}}}}
\end{equation}
From (\ref{w_k}), the phantom behavior of K-essence scalar field
($w_{K}<-1$) can be achieved when the parameter $\chi$ lies in the
interval $1/3<\chi<1/2$. Using $\dot{\phi}^2=2\chi$ and changing the
time derivative to the derivative with respect to $x=\ln{a}$, we
have
\begin{equation}\label{primephi}
\phi^{\prime}=\frac{1}{H}\sqrt{\frac{4+2\frac{a^{\frac{3}{n}}}{c-a^{\frac{3}{n}}}}
{4+3\frac{a^{\frac{3}{n}}}{c-a^{\frac{3}{n}}}}}
\end{equation}
The integration of (\ref{primephi}) yields
\begin{equation}\label{dynam77}
\phi(a)-\phi(a_0)=\int^{a}_{0}\frac{1}{aH(a)}\sqrt{\frac{4+2\frac{a^{\frac{3}{n}}}{c-a^{\frac{3}{n}}}}
{4+3\frac{a^{\frac{3}{n}}}{c-a^{\frac{3}{n}}}}}da
\end{equation}
The dynamics of reconstructed K-essence field via the evolutionary
form of polytropic dark energy is given by (\ref{dynam77}). The
K-essence polytropic gas model can explain the accelerating universe
and also behaves as a phantom model provided $1/3<\chi<1/2$. Same as
previous section we use the numerical method to calculate the
reconstructed dynamics. In figure (6) the evolution of reconstructed
K-essence is plotted as a function of cosmological redshift
$z=1/a-1$ for different values of the model parameters $n$ and $c$.
Here we choose $\phi(z=0)=0$. The scalar field decreases from up to
zero at the present time. In left panel we find a faster rate of
evolution when $n$ decreases. In right panel we see the faster
evolution of dynamics of reconstructed K-essence field for higher
values of model parameter $c$.

\begin{center}
\begin{figure}[!htb]
\includegraphics[width=7.6cm]{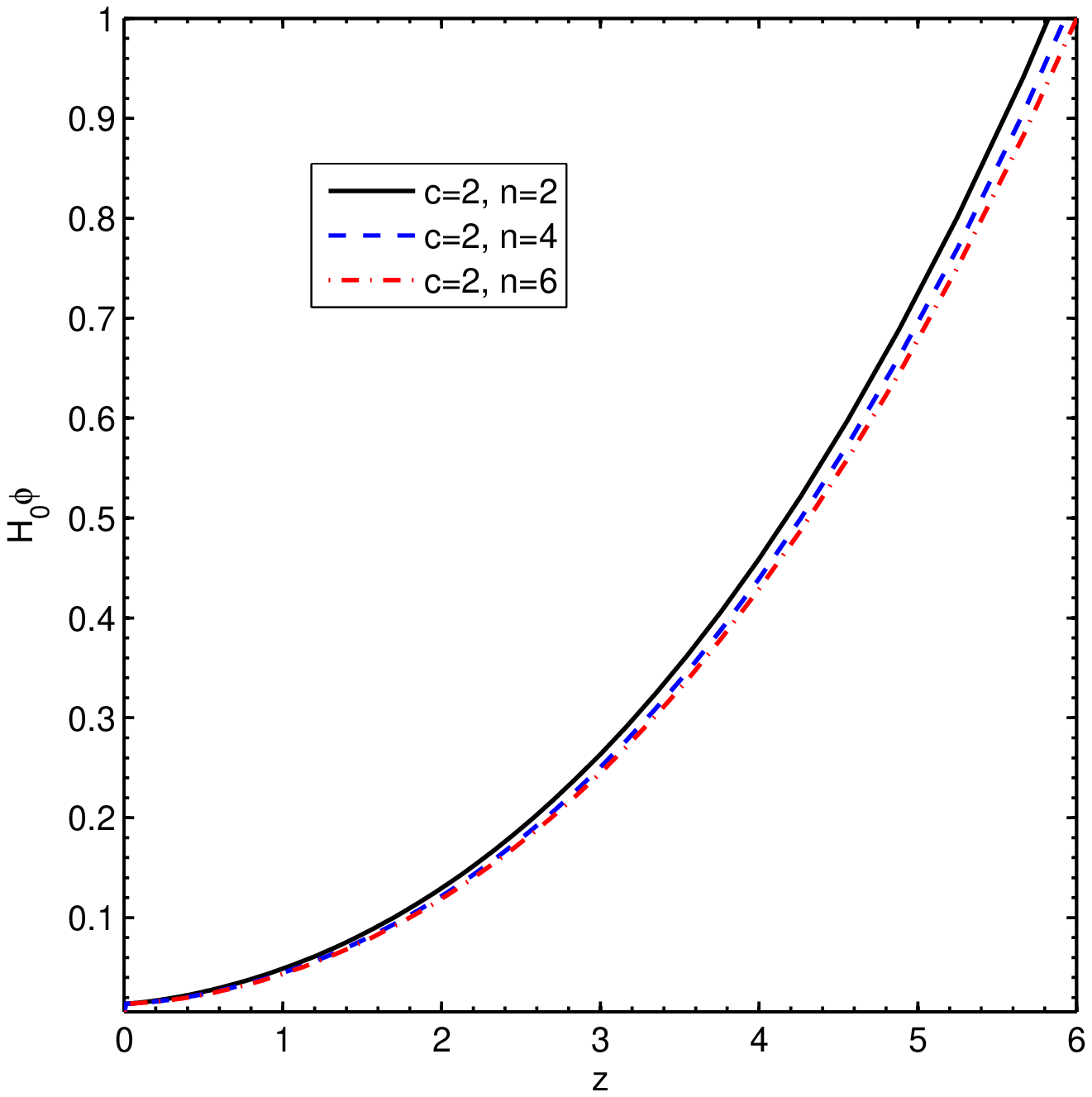}\includegraphics[width=7.6cm]{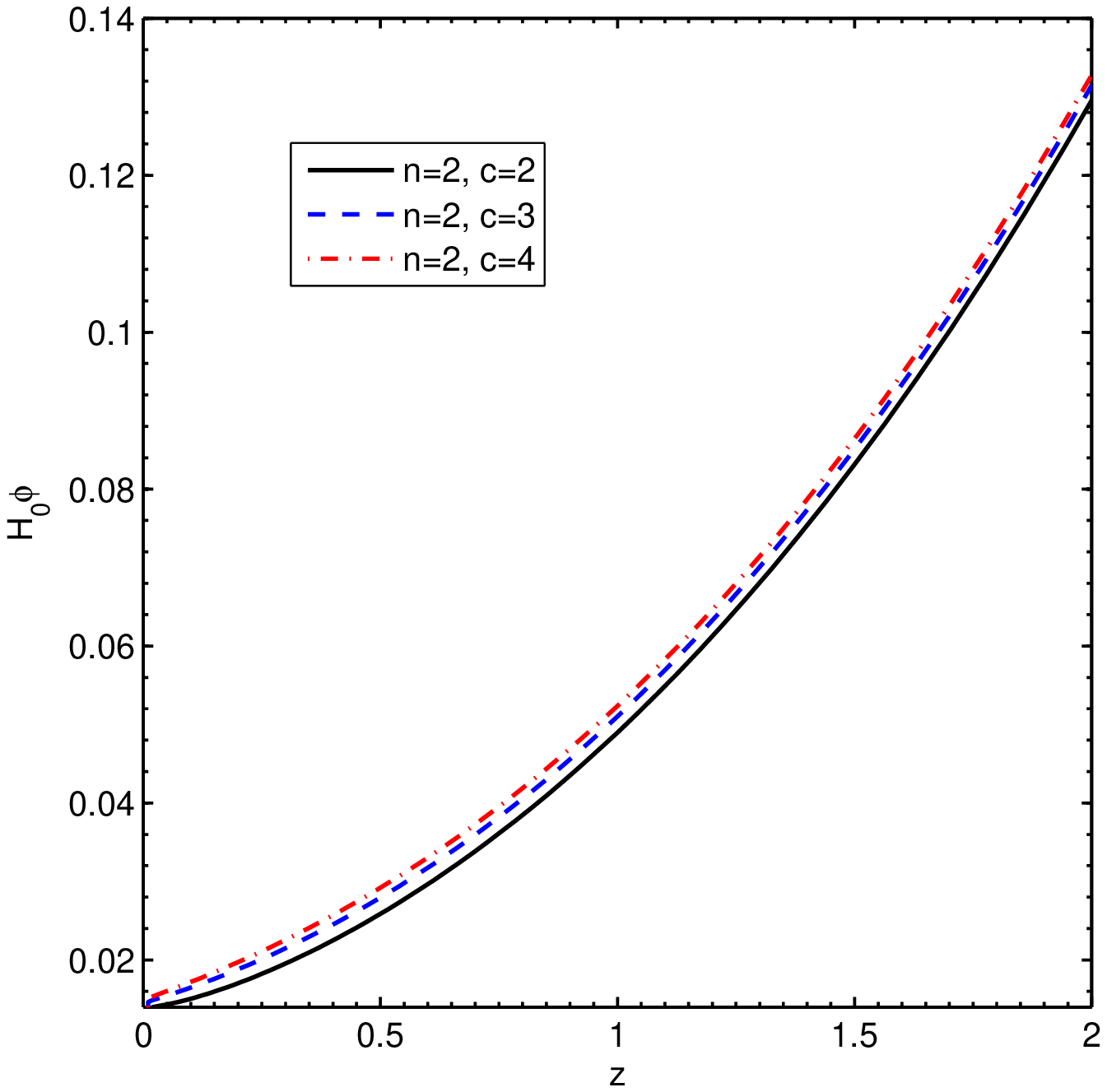}
\caption{The evolution of scalar field $\phi$ in terms of
cosmological redshift parameter $z$ for different values of $n$ and
$c$ as described in legend.}
\end{figure}
\end{center}

\section{Conclusion}
In summary, we considered the FRW cosmology with polytropic gas
model of dark energy. The polytropic gas model can explain the
cosmic acceleration of the universe and also behaves as a phantom or
quintessence dark energy models, depending on the model parameters.
One of the benefits of polytropic gas model is that it can cross the
phantom line without a need to interaction between dark energy and
dark matter. However, this model same as other phenomenological
models of dark energy, suffers from the singularity. This
singularity tacks place at $a_s=c^{n/3}$.\\
We also suggested a polytropic gas model of tachyon and K-essence
scalar field models. We adopt the viewpoint of that the scalar field
models of dark energy are effective theories of an underlying theory
of dark energy. We established a connection between the scalar field
models including tachyon and K-essence energy densities and the
polytropic gas dark energy model. We reconstructed the potential and
the dynamics of these scalar fields, numerically, according to the
evolutionary form of polytropic gas dark energy model. The
reconstructed scalar fields increases with redshift $z$ but. In an
other words, they decreases as the universe expands. This behavior
of reconstructed scalar fields via the evolution of polytropic gas
model is similar with other forms of dark energy models such as
tachyon reconstructed of new agegraphic model \cite{tach-age},
tachyon, dilaton and quintessence reconstructed of holographic dark
energy model \cite{tach-hol}.

\end{document}